# ANALYSIS OF THE NON-LTE LITHIUM ABUNDANCE FOR A LARGE SAMPLE OF F-, G-, AND K-GIANTS AND SUPERGIANTS

L. S. Lyubimkov and D. V. Petrov

*A five-dimensional interpolation method and corresponding computer program are developed for using published calculations to determine the non-LTE correction $\Delta_{NLTE}$ to the lithium abundance $\log\varepsilon(Li)$ derived from the Li I 6707.8 Å line. The $\Delta_{NLTE}$ values is determined from the following five parameters: the effective temperature $T_{eff}$, the acceleration of gravity $\log g$, the metallicity index [Fe/H], the microturbulent velocity $V_t$, and the LTE Li abundance $\log\varepsilon(Li)$. The program is used to calculate values of $\Delta_{NLTE}$ and the non-LTE Li abundance for 91 single bright giants from the list of Lebre, et al.*

*By combining these results with data for 55 stars from the previous paper, we obtain the non-LTE values of $\log\varepsilon(Li)$ for 146 FGK-giants and supergiants. We confirm that, because of the absence of the Li line in the spectra of most of these stars, it is only possible to estimate for them an upper limit for the Li abundance. A large spread is confirmed in $\log\varepsilon(Li)$ for stars with masses $M \leq 6M_\odot$. A comparison of these results with model calculations of stars confirms the unique sensitivity of the lithium abundance to the initial rotation velocity $V_0$. We discuss the giants and supergiants with lithium abundances $\log\varepsilon(Li) = 1.4 \pm 0.3$, which could have a rotation velocity $V_0=0$ km/s and have already undergone deep convective mixing. Li-rich giants with lithium abundances $\log\varepsilon(Li) \geq 2$ and nearly up to the initial value of $\log\varepsilon(Li) = 3.2 \pm 0.1$ are examined. It is shown that the fraction of Li-rich giants with $V_0 \approx 0 - 50$ km/s is consistent with current evolu-tionary models. The other stars of this type, as well as all of the "super Li-rich" giants, for which the standard theory is untenable, can be explained by invoking the hypothesis of recent lithium synthesis in the star or an alternative hypothesis according to which a giant planet is engulfed by the star.*

## 1. Introduction

Lithium, the third element in the periodic table, occupies a special place among the light elements. Of these, it is certainly the most sensitive indicator of stellar evolution. Current data on the abundance of lithium in the atmospheres of stars of different types in different stages of evolution have been reviewed by Lyubimkov [1]. There it was pointed out that the observed abundance of this element is often in conflict with theoretical predictions.

Current model calculations of stars with rotation show that even in the first stage of evolution, the Main sequence (MS) stage where hydrogen is burning in the nucleus of a star, the amounts of some light elements in the star's atmosphere can vary because of mixing induced by the rotation. In particular, the lithium abundance can be sharply reduced if the initial rotation velocity $V_0$ is high enough, at $V_0 > 50$ km/s [1]. Thus, $V_0$ is just as important a parameter in studies of the evolution of the abundance of lithium in stars as the stellar mass M. We find that, in terms of its sensitivity to $V_0$, lithium can be regarded as unique (see below).

In the MS stage, stars with masses M from 3 to $20 M_\odot$ are seen as early and medium B-stars. Because of the high effective temperatures of these stars ($T_{eff} = 15000$-$30000$ K) and the low ionization potential of Li I, lithium lines, in particular the strongest Li I resonance line at 6707.8 Å, are not observed in the spectra of these stars. This line can only be observed in comparatively cold stars with effective temperatures $T_{eff} < 8500$ K. A-, F-, G-, and K-giants and supergiants are of particular interest among these stars; they are in the evolutionary stage subsequent to the MS stage of stars with the same masses $M = 3 - 20 M_\odot$. In this stage, deep convective mixing (DCM) takes place; it can lead to additional changes in the amounts of light elements (including lithium) produced earlier in the MS stage.

It has been shown [2] in a study of 55 F- and G-supergiants and bright giants in the galaxy that many of these stars exhibit no lithium in their spectra, while the amount of lithium in other stars varies over rather wide ranges. This behavior can be explained on the whole by current model calculations of rotating stars. Nevertheless, some problems have not been solved. Thus, it is important to extend the list of FGK-giants and supergiants for which reliable estimates of the lithium content $\log\varepsilon(\text{Li})$ have been obtained.

It should be emphasized in connection with the reliability of the measured lithium abundances $\log\varepsilon(\text{Li})$ that calculations of the Li I lines, including the resonance 6707.8 Å line, generally require abandonment of the assumption of LTE (local thermodynamic equilibrium); that is, a non-LTE analysis is required. In particular, the non-LTE corrections to $\log\varepsilon(\text{Li})$ for young giants and supergiants can be has high as 0.4 dex [2]. The calculations of Lind, et al. [3], show that these corrections depend on such parameters of a star as the effective temperature $T_{eff}$, the acceleration of gravity $\log g$, the metallicity index [Fe/H], the microturbulence velocity $V_t$, and the value of $\log\varepsilon(\text{Li})$ itself.

As an illustration, Fig. 1 [3] shows the non-LTE correction $\Delta_{\text{NLTE}}$ for the Li I 6707.8 Å line as a function of $T_{eff}$ with $\log g = 2.0$ for four values of $\log\varepsilon(\text{Li})$. Here [Fe/H]=0 and $V_t$=2 km/s. This figure gives a general idea of the range of values of $\Delta_{\text{NLTE}}$ to be expected in the FGK-giants and supergiants discussed in the following. It can be seen that for temperatures $T_{eff} \geq 6000$ K the correction $\Delta_{\text{NLTE}}$ is small for all abundances $\log\varepsilon(\text{Li})$, while for $T_{eff} \leq 5000$ K, the correction is substantial and depends significantly on $\log\varepsilon(\text{Li})$.

Two problems are posed in this paper. Since non-LTE calculations of the Li 6707.8 Å line for specific stars are rather time consuming, our first problem was to simplify these calculations: this was done by creating a computer program for finding the non-LTE corrections $\Delta_{NLTE}$ to the found LTE abundance of lithium log$\varepsilon$(Li) by interpolating a grid of values of $\Delta_{NLTE}$ calculated in Ref. 3. Our second problem involved using this program to determine the non-LTE abundances of lithium for a large group of FGK-giants with known LTE abundances of lithium log$\varepsilon$(Li). On combining these results with data from Ref. 2 and, thereby, substantially extending the list of stars of this type with known non-LTE estimates for log$\varepsilon$(Li), we were able to verify some of the major conclusions of Ref. 2. Here log$\varepsilon$(Li) is given on a standard logarithmic scale on which log$\varepsilon$(H) = 12.00 for hydrogen.

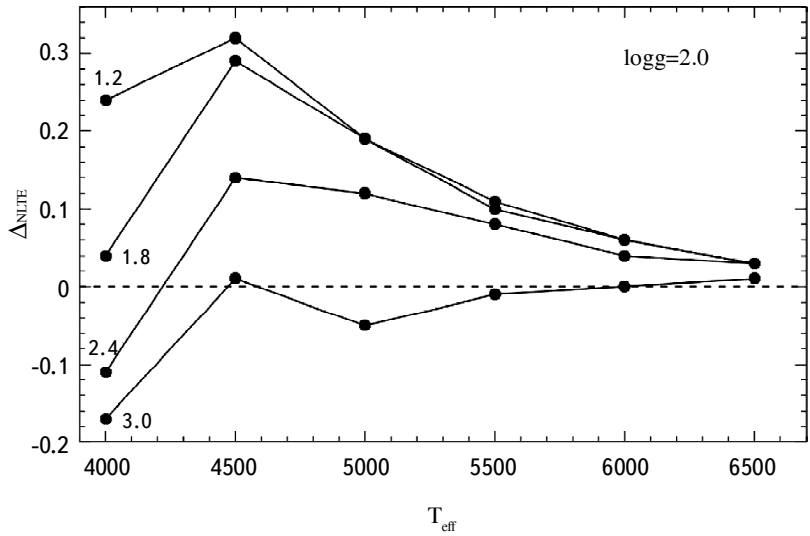

Fig. 1. The non-LTE correction $\Delta_{NLTE}$ to the lithium abundance log $\varepsilon$(Li) based on the Li I 6707.8 Å line as a function of the effective temperature $T_{eff}$, calculated for log$g$ = 2 and log $\varepsilon$(Li) = 1.2, 1.8, 2.4, and 3.0 (data from Ref. 3).

## 2. Five-dimensional interpolation of non-LTE corrections to log $\varepsilon$(Li)

Lind, et al. [3], have calculated the non-LTE corrections $\Delta_{NLTE}$ to the LTE values of log$\varepsilon$(Li) for two Li I lines: the resonance 6707.8 Å line and the subordinate 6103.6 Å line. These corrections depend on five parameters: the effective temperature $T_{eff}$, the acceleration of gravity log$g$, the metallicity index [$Fe/H$], the microturbulence velocity $V_t$, and the LTE abundance log$\varepsilon$(Li) of Li. The calculations were done for the following ranges of these parameters: $T_{eff}$ from 4000 to 8000 K; log$g$ from 1.0 to 5.0; [$Fe/H$] from -3.0 to 0.0; $V_t$ from 1.0 to 5.0 km/s; and log$\varepsilon$(Li) from -0.3 to 4.2. Some individual cases of applying the corrections of Ref. 3 for estimating log$\varepsilon$(Li) based on the 6707.8 Å line for a few stars exist in the literature. But, for correct use of the data from Ref. 3, especially for non-LTE analysis of log$\varepsilon$(Li) for a large group of stars, it is desirable to have a program which can be used to determine a value of $\Delta_{NLTE}$ immediately for each star based on its parameters $T_{eff}$, $\log g$, $[Fe/H]$, $V_t$, and $\log\varepsilon(Li)$. A program of this type, LiCOR, has been developed by one of the authors (D.V.P.).

Methods of linear interpolation (for a single variable) are well known. Explicit formulas for bilinear (with respect to two parameters) and trilinear (with respect to three parameters) can also be found in the literature. Our problem involved interpolation with respect to the five parameters listed above. We used the standard idea of multidimensional interpolation that is discussed, for example, in section 20.5-6 of the handbook of Korn and Korn [4]. It involves successive linear interpolation with respect to each of the parameters with the others kept constant. We do not give the interpolation formulas that we derived here because of their complexity. In order to check the accuracy of the program LiCOR that is based on these formulas, we applied the program to stars that had been examined in Ref. 2. We note that in that paper, the lithium abundance $\log\varepsilon(Li)$ was determined from the Li I 6707.8 Å line.

For many of the stars in Ref. 2 it was possible to estimate only an upper bound for $\log\varepsilon(Li)$ because the Li 6707.8 Å line does not appear in their spectra. Exact values of LTE and non-LTE $\log\varepsilon(Li)$ were, however, obtained for 11 supergiants and giants of classes F5 to K0 (see Table 1 of Ref. 2). Thus, the non-LTE corrections $\Delta_{NLTE}$ obtained in Ref. 2 using direct non-LTE calculations are known for these 11 stars. It is interesting to compare the values of $\Delta_{NLTE}$ found for these stars in Ref. 2 with the results of calculations using the LiCOR program. In Fig. 2a the corrections $\Delta_{NLTE}$ are shown as functions of the effective temperature $T_{eff}$. There the solid circles correspond to data from Ref. 2 and the crosses, to the values of $\Delta_{NLTE}$ found for the same 11 stars using the LiCOR program (here the parameters $T_{eff}$, $\log g$, $[Fe/H]$, $V_t$, and $\log\varepsilon(Li)$ are the same as in Ref. 2).

A trend in the dependence of $\Delta_{NLTE}$ on $T_{eff}$ can be seen in Fig. 2a: when $T_{eff}$ is lower, the correction $\Delta_{NLTE}$ is larger on the average. On the whole, the corrections $\Delta_{NLTE}$ found using the program LiCOR and using the non-LTE calculations of Ref. 2 are in good agreement in this figure. For most of these stars (9 out of 11), the difference between the values of $\Delta_{NLTE}$ obtained independently by the two methods is very small: it varies from +0.03 to -0.02 dex (averaging +0.01 dex). This small discrepancy can be completely attributed to errors in the interpolation. The difference is larger for only two of the stars: -0.09 dex for HD 17905 and +0.11 dex for HR 8313 (these are indicated in Fig. 2).

From many years of experience with non-LTE calculations for different chemical elements it is known that data from different authors on lines of the same atoms or ions can differ significantly. This is explained by differences in the methods used for the non-LTE calculations, as well as in the models for stellar atmospheres. Lind, et al. [3], compared their results for lithium with data from other authors and concluded that the difference in the non-LTE corrections is as large as 0.20 dex. In particular, it follows from Ref. 3 that, for the above reasons, the expected difference in $\Delta_{NLTE}$ based on the data from Ref. 2 and the LiCOR program applied to the data of Ref. 3 can reach 0.12 dex. Thus, even in the case of the stars HD 17905 and HR 8313, for which the largest discrepancies in $\Delta_{NLTE}$ were found (about ±0.10 dex), these discrepancies should be entirely explainable by differences in the non-LTE computational methods and in the models for the atmospheres.

In discussing the accuracy of the corrections $\Delta_{NLTE}$ which we have found using the LiCOR program, it should be noted that the data of Ref. 3 upon which this program is based were obtained using a better non-LTE method than in Ref. 2. In particular, in Ref. 3, as opposed to Ref. 2, inelastic collisions with hydrogen atoms have been taken into account. In addition, the models of the atmosphere used in Ref. 3 and calculated using an improved version of the MARCS program appear to be more accurate than the models used in Ref. 2. This indicates that the corrections $\Delta_{NLTE}$ obtained with the LiCOR program will be more accurate than the $\Delta_{NLTE}$ in Ref. 2.

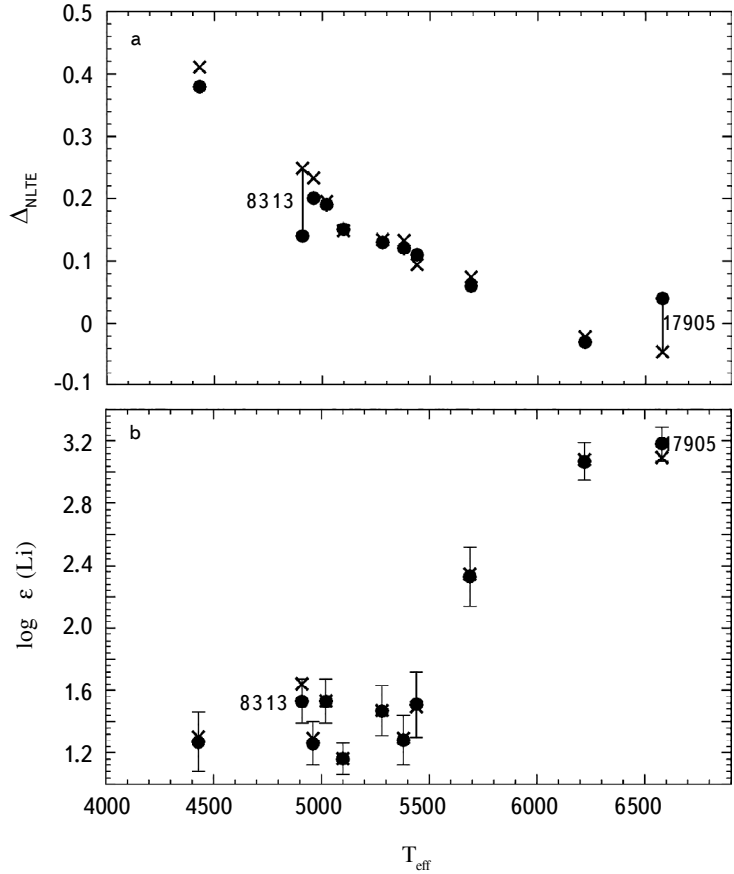

Fig. 2. Comparisons of (a) the non-LTE corrections $\Delta_{NLTE}$ and (b) the non-LTE lithium abundances $\log \varepsilon(Li)$ for 11 supergiants and giants based on data from two sources: Ref. 2 (solid circles) and the present paper (crosses).

Figure 2b shows that if the dependence of the non-LTE abundances $\log\varepsilon(Li)$ on $T_{eff}$ are considered instead of the dependence of $\Delta_{NLTE}$ on $T_{eff}$, the difference between the two systems of $\Delta_{NLTE}$ corrections is within the limits of error for the values of $\log\varepsilon(Li)$. This conclusion holds even for the stars HD 17905 and HR 8313 (the error bars in Fig. 2b correspond to the data of Ref. 2). Thus, the calculations of $\Delta_{NLTE}$ using the LiCOR program provide quite sufficient accuracy for studies of the non-LTE abundance of Li in a sufficiently large group of stars.

We have also done test calculations with the LiCOR program for the cold K-giant HD 77361, which is of interest because it is one of the rare "super Li-rich giants." The following parameters have been found for this star [5]: $T_{eff}=4370$ K, $\log g=2.30$, $[Fe/H]=-0.01$, $V_t=1.1$ km/s, and an LTE Li abundance of $\log\varepsilon(Li)=3.80$ (based on the 6707.8 Å line). The LiCOR program gives a non-LTE correction of $\Delta_{NLTE} = -0.19$ for these parameters; this differs by -0.08 dex from the $\Delta_{NLTE} = -0.11$ found in Ref. 5.

## 3. Determining the non-LTE values of log ε(Li) for the bright giants in the list of Lebre, et al. [6]

As noted above, a non-LTE analysis of the lithium abundance in the atmospheres of 55 F- and G-stars in luminosity classes I and II was made in Ref. 2. The abundances log ε(Li) found for these stars, as well as a subsequent comparison with theory, lead to a series of interesting conclusions. The paper by Lebre, et al. [6], opens up the possibility of greatly expanding the list of stars of this type. The LTE abundances of lithium (based on the Li I 6707.8 Å line) were determined for 145 F-, G-, and K-bright giants (luminosity class II and some II-III) in that paper. It is significant that this list [6] includes both single and spectrally-binary stars. In order to apply our program LiCOR to these stars, we first select the stars that are suitable for subsequent analysis.

We eliminated spectral-binary stars from the list of Ref. 6. We note that binary stars were also not considered in Ref. 2. It is important that a correct analysis of binary systems requires an individual determination of the parameters of both components, including $T_{eff}$, log$g$, M, and the lithium abundance log ε(Li). A study of the binary star ι Peg [7] in which the components A and B had different but high Li abundances can serve as an example of this kind of analysis. (If the small non-LTE corrections found using the LiCOR program are taken into account, log ε(Li) =3.17 and 2.59 for components A and B, respectively.) Since binary stars were analyzed as single stars in Ref. 6 and average values of some sort were obtained for these parameters, they have been eliminated from our analysis.

We also excluded the lowest mass stars with masses $M < 1M_\odot$ for which the evolutionary tracks that we have used could not provide a reliable estimate of *M* (see below). Therefore, we have retained 91 single stars from the list of Ref. 6 as suitable for further analysis.

As pointed out above, use of the LiCOR program requires knowledge of the following input parameters: $T_{eff}$, log$g$, [*Fe/H*], $V_t$, and the lithium abundance logε(Li). Values of $T_{eff}$ and log$g$ for a number of the stars in Ref. 6 were taken from the literature. When these data were lacking, the effective temperature $T_{eff}$ in Ref. 6 was determined from the photometric index (*B-V*) and the acceleration of gravity was taken to be log$g$ = 2.0. In addition, for some of these stars it was assumed that $V_t$= 2 km/s and [*Fe/H*] = 0. Thus, for some fraction of the stars in Ref. 6, we have specified the approximate parameters log$g$=2.0, $V_t$=2 km/s, and [*Fe/H*]=0.0. It is necessary to evaluate how this could influence the resultant lithium abundance.

As for the metallicity index [*Fe/H*], it can be said that for most of the stars in Ref. 6 this parameter is actually close to zero; it ranges from +0.5 to -0.5 dex (see Table 1 of Ref. 6). A value of $V_t$ = 2 km/s for the microturbulence parameter may be somewhat low for the bright giants (luminosity class II). According to an accurate determination of the fundamental parameters for 63 supergiants and bright giants in the galaxy [8] (these data were used in Ref. 2), $V_t$ for stars in luminosity class II mostly lies within a range of $V_t \approx 2.5 - 3.5$ km/s. However, our estimates show that a possible low value of $V_t$ in Ref. 6 should not have a significant influence on the determination of logε(Li).

As for the value $\log g = 2.0$, it corresponds fairly well to F- and G-stars in luminosity class II. According to this same reference [8], they typically have $\log g = 2.0$-$2.5$. Fortunately, the lithium abundance $\log\varepsilon(\text{Li})$ determined here is insensitive to errors in $\log g$, so replacing the exact value of $\log g$ by $\log g = 2.0$ could not have a significant effect on the lithium abundance found in Ref. 6. The very low sensitivity of the lithium abundance to $\log g$ is confirmed, for example, by the calculations of Ref. 3, which imply that changing $\log g$ by 1.0 dex (from 2.0 to 3.0) changes the equivalent width $W$ of the Li I 6707.8 Å line by only a few percent. The corresponding change in $\log\varepsilon(\text{Li})$ is only 0.02-0.03 dex. $W$ is much more sensitive to variations in $T_{eff}$. Changing $T_{eff}$ by 100 K can change $W$ by 50-60%.

The parameters $T_{eff}$ and $\log g$ from Ref. 6 are not only needed for determining the corrections $\Delta_{NLTE}$ to the lithium abundance for the 91 stars we have selected from Ref. 6, but also for determining the masses $M$ of these stars. We used the same evolutionary tracks [9] as in Refs. 2 and 8 for estimating $M$. Figure 3 shows the evolutionary diagram in the $T_{eff}$- $\log g$ plane; besides the positions of the stars that we have studied (crosses), it shows a number of tracks [9] calculated for masses $M$ from 0.8 to $16\,M_\odot$ (here $M_\odot$ is the mass of the sun). As mentioned above, these tracks [9] can only be used to estimate $M$ for stars with $M \geq 1 M_\odot$.

We have used the program LiCOR to find the non-LTE corrections $\Delta_{NLTE}$ for the 91 selected stars. In Fig. 4 the resulting values of $\Delta_{NLTE}$ are plotted as a function of effective temperature $T_{eff}$. Here a trend in the variation of $\Delta_{NLTE}$ with decreasing $T_{eff}$ can be seen clearly: while the correction $\Delta_{NLTE}$ is close to zero for $T_{eff} = 6000$-$7000$ K, it increases with falling $T_{eff}$ and reaches 0.3-0.4 dex for the coldest stars with $T_{eff} = 4100$-$4500$ K. The existence of this trend was pointed out during the discussion of Fig. 2a.

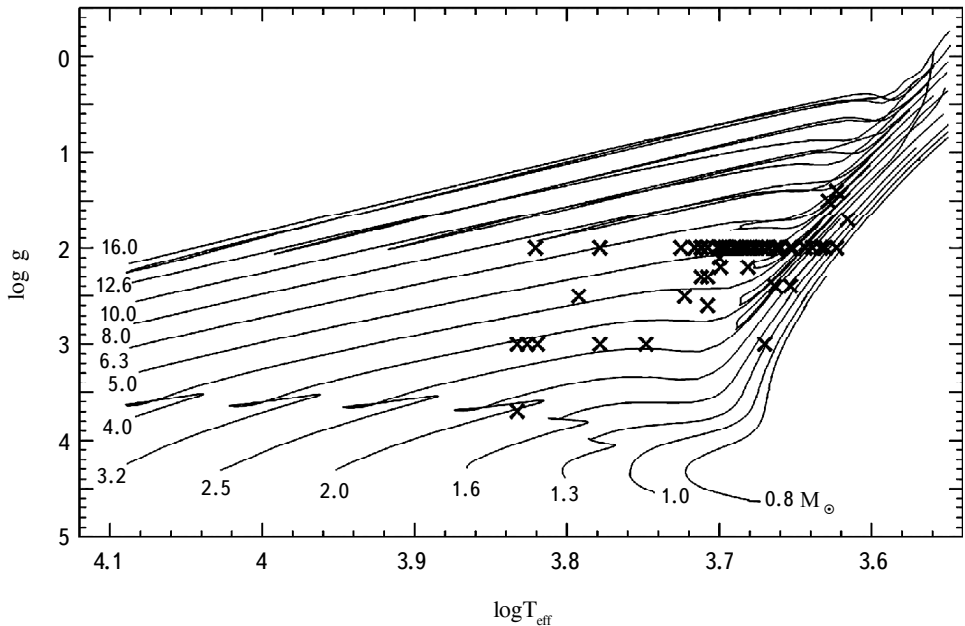

Fig. 3. Positions of the 91 stars selected from the list of Lebre, et al. [6], in the $T_{eff}$- $\log g$ plane. A number of evolutionary tracks [9] are shown.

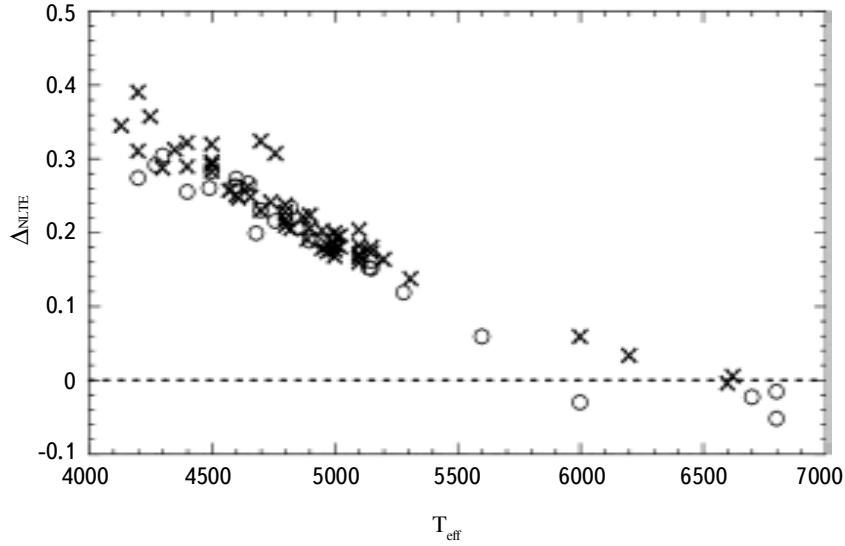

Fig. 4. The corrections $\Delta_{\text{NLTE}}$ calculated for the 91 stars from the list of Ref. 6 as functions of effective temperature $T_{\text{eff}}$. The open circles correspond to exact values of log ε(Li) and the crosses, to an upper bound on log ε(Li).

In comparing this figure with Fig. 1, it might be asked why no spread in $\Delta_{\text{NLTE}}$ is observed at low $T_{\text{eff}}$ owing to differences in the lithium abundance logε(Li). The answer is that all the stars with $T_{\text{eff}}$ ranging between 4000 and 5500 K that were selected from Ref. 6 have low LTE abundances logε(Li) < 1.5 which, according to Fig. 1, correspond to comparatively high values of $\Delta_{\text{NLTE}}$.

We used the resulting corrections $\Delta_{\text{NLTE}}$ to determine the non-LTE abundance log ε(Li) of lithium for the 91 selected stars. As in Ref. 1, we should again note the important fact that the Li I 6707.8 Å is absent in the spectra of most FGK-giants and supergiants, so for these stars it is only possible to estimate an upper bound on the lithium abundance log ε(Li). In particular, of the 91 objects shown in Fig. 4, for 63 (69%) of these stars the data of Ref. 6 can only provide an upper bound on the lithium abundance (crosses), while an exact estimate of log ε(Li) was obtained for 28 (31%) of the stars (open circles).

## 4. Analysis of the non-LTE lithium abundances for the 146 FGK-giants and supergiants

As a whole, the above non-LTE Li abundances for 91 FGK-stars in luminosity classes II and II-III, together with the Li abundances from Ref. 2 for 55 stars in luminosity classes I, II, and (partially) III, provide a quite interesting basis for analyzing the lithium content of 146 FGK-giants and supergiants. In Fig. 5 the non-LTE log ε(Li) for these stars is plotted as a function of mass *M*, which is the most important parameter from the standpoint of stellar evolution. In the case of lithium, as mentioned in the *Introduction*, the initial rotation velocity may be an equally important parameter. (Note that Fig. 5 is an expanded version of Fig. 7 in Ref. 2.)

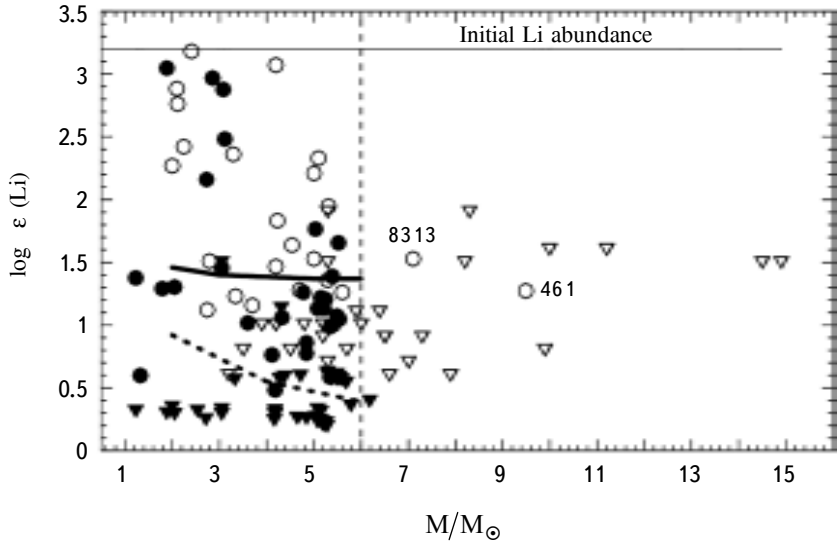

Fig. 5. Non-LTE lithium abundance as a function of mass. The solid symbols correspond to data for the 91 stars obtained in the present paper and the hollow symbols, to data for 55 stars from Ref. 2. The circles are exact values of log ε(Li) and the triangles are upper bounds on log ε(Li). The smooth curve is the theoretical prediction for cold giants with $V_0$ = 0 km/s after the DCM and the dashed curve is the same for giants with $V_0$ = 50 km/s.

One of the main results of the earlier paper [2] was that these observations and stellar model calculations show that the behavior of $\log\varepsilon(Li)$ differs fundamentally for stars with masses of $M \leq 6 M_\odot$ and $M > 6 M_\odot$. For the 91 stars discussed in this paper, we obtained masses ranging from 1.2 to $6.2 M_\odot$. Thus, almost all the stars indicated by the solid symbols (circles and triangles) in Fig. 5 lie within the region $M \leq 6 M_\odot$. As for the stars in the list of Ref. 2 (open circles and triangles in Fig. 5), they have substantial variations in $\log\varepsilon(Li)$, ranging from values close to the initial lithium abundance $\log\varepsilon(Li) = 3.2 \pm 0.1$ [1] to complete absence of any evidence of lithium in the spectra of many of the stars. As noted above, for 63 of the 91 stars it was possible only to estimate a upper bound for $\log\varepsilon(Li)$, while an exact value of $\log\varepsilon(Li)$ was found for 28 of the stars.

The stars with a high lithium abundances $\log\varepsilon(Li) \geq 2.0$ in Fig. 5, which belong to the "Li-rich" type, are of special interest. Stars of this type are known to form a small fraction of all FGK-giants and supergiants [1] and this situation leads to the assumption that this phase of evolution is of short duration. We note an important fact discovered in Ref. 2: all Li-rich giants have masses $M < 6 M_\odot$.

Of the 91 stars we have studied, five are of the "Li-rich" type. Their parameters are listed in Table 1. We used the parallaxes p from the SIMBAD data base (http://simbad.u-strasbg.fr/simbad/sim-fid) to find their distances $d = 1/\pi$. In Table 1, d ranges from 39 to 160 pc; i.e., these stars are fairly close. The parameters $T_{eff}$, log$g$, and $V$sin$i$ in Table 1 are taken from Lebre, et al. [6]. We determined the masses $M$ and non-LTE abundances $\log\varepsilon(Li)$ in Table 1.

TABLE 1. Parameters of Five "Li-rich" from the List of Lebre, et al. [6]

| HD | $m_V$, mag | $\pi$, mas | d, pc | $T_{eff}$/log$g$ [6] | $V\sin i$, km/s | $M/M_\odot$ | log$\varepsilon$(Li) |
|---|---|---|---|---|---|---|---|
| 65228  | 4.20 | 6.25±0.23  | 160 | 5600/3.0 | 14 | 2.7 | 2.16 |
| 186155 | 5.07 | 20.15±0.16 | 50  | 6000/3.0 | 59 | 2.9 | 2.97 |
| 204509 | 6.63 | 9.52±0.91  | 105 | 6700/3.0 | 19 | 3.1 | 2.88 |
| 216756 | 5.91 | 25.63±0.34 | 39  | 6800/3.7 | 12 | 1.9 | 3.05 |
| 218043 | 6.78 | 15.51±0.45 | 64  | 6800/3.0 | 20 | 3.1 | 2.49 |

It is interesting that one of these giants, HD 65228 (HR 3102), was studied in Ref. 2. In Table 2 we compare the parameters of this star according to the data of Refs. 2 and 6. It can be seen that even with an apparently good agreement in the values of $T_{eff}$ (a difference of 90 K), there is a significant discrepancy of about 0.8 dex in the values of log$g$. As noted above, the lithium abundance log $\varepsilon$(Li) is sensitive to variations in $T_{eff}$, but depends weakly on log$g$. Our estimates show that a difference of 0.8 dex in log$g$ can change log $\varepsilon$(Li) by only 0.02 dex. According to Table 2, the abundance log $\varepsilon$(Li) found in Ref. 2 differs from that found in this paper by 0.17 dex. (It is interesting that the non-LTE correction is the same in both cases: $\Delta_{NLTE}$ = 0.06 dex.) Our estimate shows that this discrepancy is explained mainly by the above noted difference of 90 K in the values of $T_{eff}$.

The example of the giant HD 65228 is also interesting in that the following anomalies in the abundances of carbon and nitrogen (relative to the solar abundances) were found: a deficit of carbon with [C/Fe] = –0.24, an excess of nitrogen with [N/Fe] = 0.41, and an elevated ratio [N/C] = 0.65 [10]. Anomalies of this kind in C and N are typical of giants that have passed through the deep convective mixing (DCM) phase, but in this case, all the lithium should

TABLE 2. A Comparison of the Parameters of the Star HD 65228 (HR 3102) Based on Data from Two References

| $T_{eff}$ | log$g$ | [Fe/H] | $V\sin i$, km/s | log$\varepsilon$(Li) LTE | log$\varepsilon$(Li) non-LTE | References |
|---|---|---|---|---|---|---|
| 5690 | 2.17 | 0.11 | 12 | 2.27 | 2.33 | [2] |
| 5600 | 3.00 | 0.00 | 14 | 2.10 | 2.16* | [6] |

* The non-LTE value of log $\varepsilon$(Li) found in this paper.

have been burnt up. The high value of log ε(Li) = 2.3 for HD 65228, together with the anomalies in C and N, suggests recent synthesis of lithium (see below).

As shown before [2], stellar model calculations without rotation ($V_0$ = 0 km/s) give an atmospheric lithium abundance for FGK-supergiants with masses $M = 2 - 6 M_\odot$ which have passed through the DCM phase that is close on the average to log ε(Li) = 1.4 (the smooth curve in Fig. 5). Objects of this sort in Fig. 5 can be stars lying within the range log ε(Li)=1.4 ±0.3 , i.e., with log ε(Li)=1.1–1.7 (including the possible errors log ε(Li). It can be seen that for objects with masses $M \leq 6 M_\odot$, besides the 10 stars from Ref. 5 (open circles) this range includes another 11 stars of the newly studied set (solid circles). There is a total of 21 stars, which represents a significant fraction (14%) of the total number of stars (146).

The determination of the non-LTE Li abundance carried out in the present paper for a group of 91 FGK-giants did not add any new (compared to Ref. 2) data for objects with masses $M > 6 M_\odot$. Thus, we only summarize briefly the main conclusions of Ref. 2 regarding these objects. According to the theory, stars with $M > 6 M_\odot$ should arrive in the FGK-giant (supergiant) state with very low Li abundances that cannot be detected. This applies even to stars without rotation ($V_0$ = 0 km/s) for which the Li abundance falls off sharply immediately after emergence from the MS stage. The data of Ref. 2 and Fig. 5 mostly confirm this prediction: lithium does not show up in the spectra of most stars with $M > 6 M_\odot$. The two cold supergiants HR 461 (K0 Ia) and HR 8313 (G5 Ib) (indicated in Fig. 5), which merit special attention, are an exception.

These two stars can be explained in two ways. On one hand, given their low temperatures $T_{eff}$ (4430 and 4910 K), it can be assumed that they have already passed through the DCM phase. It is noteworthy that the lithium abundances found for them (1.3 and 1.5, respectively) are very close to the value $\log \varepsilon(Li) \approx 1.4$ predicted for models with $V_0$ = 0 km/s and masses $M \leq 6 M_\odot$ (see above). It seems possible that the theoretical predictions for $M > 6 M_\odot$ are not entirely correct. On the other hand, if the calculations for $M > 6 M_\odot$ are true, it could be that lithium synthesis has recently occurred in these stars.

## 5. The initial rotation velocity $V_0$ and the abundance of lithium

Stellar model calculations with rotation show that the amount of lithium is very sensitive to the initial rotation velocity $V_0$. The following question arises in connection with this: what are the actually observed initial velocities $\underline{V}_0$ in FGI-giants and supergiants? We recall [1] that the predecessors of these stars with masses $M = 3 - 20 M_\odot$ are early MS B-stars. It has been found [11] that the maximum in the distribution of observed rotation velocities $V\sin i$ for MS dwarfs in classes B0-B2 lies within a range of 0-20 km/s, although the entire range of values of $V\sin i$ extends to ~400 km/s. A similar result was obtained earlier in Ref. 12: the maximum in the distribution of $V\sin i$ for early MS B-stars lies within a range of 0-50 km/s. Thus, the observed rotation velocities of early MS B-stars, which are the predecessors of the giants and supergiants discussed here, suggest that the initial rotation velocity $V_0$ was low for most of them.

Data from a recent analysis of the N/O ratio in the atmospheres of 46 early MS B-stars [13] can also serve as confirmation of this effect. N/O can be regarded as one of the indicators of stellar evolution. It was shown [13] that for most of these stars this ratio did not vary during evolution in the MS (i.e., $[N/O] \approx 0$). It was concluded that these stars had low $V_0$. Strictly speaking, a comparison with models of rotating stars implies that they have $V_0 < 100$ km/s. Only when $V_0 = 200\text{-}300$ km/s are elevated values of $[N/O] = 0.4\text{-}0.8$ observed at the end of the MS, but there are few stars of this kind [13].

The abundances of C, N, and O, as well as their ratios (in particular, N/O), change during the process of evolution, but these changes are not as sensitive as the lithium abundance to the initial rotation velocity $V_0$. For example, as noted in Ref. 1, when $V_0 = 100$ km/s toward the end of the MS stage, the abundances of C and N in the atmosphere are essentially unchanged, while the amount of Li falls by several orders of magnitude and, therefore, becomes undetectable. Even for a velocity of $V_0 = 50$ km/s, the amount of lithium in the atmospheres of stars with masses of 2 and $4 M_\odot$ should fall by 0.6 and 1.1 dex, respectively toward the end of the MS; this corresponds to lithium abundances of $\log\varepsilon(\text{Li}) = 2.6$ and 2.1. After a transition into the cold giant stage and after the DCM, the amount of Li for $V_0 = 50$ km/s becomes so small ($\log\varepsilon(\text{Li}) < 1$) that it is hard to detect (see the dashed curve in Fig. 5). Lithium undoubtedly manifests a unique sensitivity to $V_0$.

It is clear from these remarks that the case of $V_0 = 0$ km/s for stars of this type is of special interest. It is possible with $V_0 = 0$ km/s that an FGK-supergiant (giant) with $M = 2-6 M_\odot$ has still not passed through the DCM phase. Stellar model calculations show that the condition $T_{eff} > 5900$ K must be met for this to be so. In this case, although it is in a rather advanced stage of evolution, a star must retain its initial lithium abundance $\log\varepsilon(\text{Li}) = 3.2 \pm 0.1$.

Stars with $\log\varepsilon(\text{Li})$ close to 3.2 are actually observed in Fig. 5. Three stars of this kind are listed in Table 1: HD 186155, 204509, and 216756, for which $\log\varepsilon(\text{Li}) = 2.88-3.05$. Two stars, HD 17905 and HR 7008, with abundances $\log\varepsilon(\text{Li}) = 3.18$ and 3.07 that are especially close to the initial $\log\varepsilon(\text{Li}) = 3.2$, have been studied in Ref. 2. Another two stars, HR 1298 and HR 2936, with $\log\varepsilon(\text{Li}) = 2.76$ and 2.88, were also studied there. All of these stars are F-giants with effective temperatures $T_{eff} \geq 6000$ K. Since $T_{eff} > 5900$ K for them, it is entirely possible that they have not yet reached the DCM phase (the alternative is a return into a region of elevated temperatures $T_{eff}$ along the "red-blue-red" loop [2]). If they began their evolution with a low rotation velocity $V_0 \approx 0$ km/s, then theoretically they should retain the initial lithium abundance in their atmosphere.

The case of $V_0 = 0$ km/s also applies directly to giants which have already passed through the DCM phase. As noted above, for these stars the theory predicts a lithium content of $\log\varepsilon(\text{Li}) \approx 1.4$ We note that Fig. 5 contains 21 stars in the group $M \leq 6 M_\odot$ with lithium abundances $\log\varepsilon(\text{Li}) = 1.4 \pm 0.3$. It is possible that the same scenario applies to two supergiants from the group with $M > 6 M_\odot$, HR 461 and HR 8313, since their lithium abundances are also close to $\log\varepsilon(\text{Li}) = 1.4$ (see above).

## 6. On the origin of lithium-rich giants

As pointed out above, cold giants and supergiants that are rich in lithium with $\log\varepsilon(\text{Li}) \geq 2$ (lithium-rich giants) are of special interest. It has been stated in the literature that Li-rich giants are an enigma for the standard theory of stellar evolution. It can, however, be assumed that some of these stars can, nevertheless, be explained in terms of current stellar model calculations.

It follows from the previous section that the case of $V_0 \approx 0$ km/s occurs fairly often for FGK-giants and supergiants. As noted above, if stars with $V_0 \approx 0$ km/s have not yet reached the DCM phase, they can be observed as Li-rich giants with lithium contents up to the initial $\log\varepsilon(Li) = 3.2$. This scenario for the appearance of Li-rich giants can operate for $V_0$ within the range from 0 to 50 km/s (see above). Thus, current models of stellar evolution can explain, in part, the existence of Li-rich giants if they had low initial rotation velocities $V_0 \approx 0-50$ km/s. In other words, some of the Li-rich giants can be fully explained in terms of the current theory of stellar evolution.

On the other hand, the theory is untenable for stars with $V_0 \approx 0-50$ km/s which have already passed through the DCM phase (the Li abundance after DCM is too low; cf. the solid and dashed curves in Fig. 5). The theory also cannot explain the origin of Li-rich giants if $V_0 \sim 100$ km/s or more. Here it should be noted that many Li-rich giants have actually passed through the DCM phase: this is indicated by the low carbon isotope ratio $^{12}C/^{13}C$ in these stars [1]. We again recall the example of the giant HD 65228 (HR 3102) from Table 1, for which anomalies in the C and N abundances were found which indicate that this lithium rich star has already passed through the DCM phase, during which all the lithium should have been burnt up.

The standard theory is completely unable to explain the origin of giants that are super-rich in lithium (the "super Li-rich giants") with Li abundances greatly in excess of the initial $\log\varepsilon(Li) = 3.2$. The cold giant HD 77361 (HR 3597) noted above is an example of a super Li-rich giant. It has a high lithium abundance $\log\varepsilon(Li) = 3.75 \pm 0.11$ determined from three, rather than one, lithium line [5]: the strong 6707.8 Å resonance line and the fainter subordinate lines at 6103.6 and 8126.4 Å. This star has a low observed carbon isotope ratio $^{12}C/^{13}C$ = 4 [14] (recall that $^{12}C/^{13}C$ = 89 for the sun), which indicates that the star has passed through the DCM phase, during which all the lithium should have been burnt up.

The untenability of the theory in all these cases forces us to seek other explanations for the high lithium abundances. Two hypotheses for explaining the phenomenon of Li-rich and super Li-rich stars are currently being discussed in the literature [1]. The first proposes recent synthesis of lithium inside a star. The second invokes an external factor: engulfment by the star of a brown dwarf or a giant planet similar to Jupiter. Besides a high Li abundance, it appears that the second hypothesis can also explain another interesting fact: unexpectedly high rotation velocities (up to 100 km/s) in a number of Li-rich stars which are utterly atypical of cold giants [1].

Both of these hypotheses are under active discussion in the current literature. Arguments are made "for" and "against" each of these hypotheses, but they have not been discussed in detail for our problem. One recent study [15] is of interest for calculations confirming the capture of a planet. The alternative hypothesis, lithium synthesis inside a star is discussed in Ref. 16 from an somewhat unexpected standpoint— the possibility of forming organic molecules in a shell surrounding an Li-rich giant.

# 7. Conclusion

Here we summarize the major results of this paper.

A five-dimensional interpolation method and a corresponding computer program LiCOR have been developed for finding the non-LTE correction $\Delta_{NLTE}$ to the lithium abundance $\log(\varepsilon Li)$ for the Li I 6707.8 Å line which is based on the calculations of Ref. 3 and uses the following five parameters: the effective temperature $T_{eff}$, the acceleration of gravity $\log g$, the metallicity index [*Fe/H*], the microturbulence velocity $V_t$, and the LTE abundance of lithium $\log \varepsilon(Li)$. A comparison with detailed non-LTE calculations for a number of stars confirms the adequately high accuracy of this method.

The LiCOR program has been used to determine $\Delta_{NLTE}$, as well as the non-LTE Li abundance, for 91 single bright giants selected from the list of Lebre, et al. [6]. On combining these results with data for 55 stars from Ref. 2, we have obtained non-LTE values of $\log \varepsilon(Li)$ for 146 FGK-giants and supergiants, which serve as a basis for testing and confirming the conclusions of Ref. 2.

The large spread in $\log \varepsilon(Li)$ for stars with masses $M \leq 6 M_\odot$ has been confirmed, beginning with values close to the initial abundance $\log \varepsilon(Li) = 3.2 \pm 0.1$ and extending to the complete absence of any signs of lithium. For most of these stars, because there is no Li in their spectra, it is only possible to obtain an upper limit of the Li abundance. As a whole, these results agree with current models of rotating stars.

We have discussed the dependence of the lithium abundance $\log \varepsilon(Li)$ on the initial rotation velocity $V_0$ and confirm that lithium manifests a unique sensitivity to $V_0$. It was pointed out that these stars, which are in the MS stage of early B-stars, have frequently a low rotation velocity $V_0$, which could play a decisive role in their evolution, including in the evolution of the lithium abundance.

It has been shown that the case of $V_0 \sim 0$ km/s has a direct relationship to giants and supergiants with abundances $\log \varepsilon(Li) = 1.4 \pm 0.3$, which form a marked fraction of the stars studied here. According to theory, they had $V_0 = 0$ km/s and have already undergone deep convective mixing.

Lithium-rich giants, which have lithium abundances $\log \varepsilon(Li) \geq 2$ and ranging up to values close to the initial $\log \varepsilon(Li) = 3.2 \pm 0.1$ or even higher (super Li-rich giants), have been discussed. It has been shown that some of the Li-rich giants for which $V_0 \approx 0$ km/s may be consistent with current evolutionary models. For the other Li-rich stars and all the super Li-rich stars, where the standard theory is untenable, it is necessary to draw on the hypothesis of recent lithium synthesis in the star or on an alternative hypothesis according to which either a giant planet similar to Jupiter or a brown dwarf is engulfed by the star.